\documentclass[11pt,a4paper]{article}
\usepackage[left=1in, right=1in, top=1in, bottom=1in]{geometry}
\usepackage{pgfplots}
\pgfplotsset{width=7cm,compat=1.15}
\usepackage{graphicx}
\usepackage{bm}
\newcommand*{\eqref}[1]{(\ref{#1})}
\usepackage[skip=8pt,font=scriptsize]{caption}
\usepackage{epstopdf}
\begin{document}
\begin{center}
{\large{Relativistic-kinematic approach to the hydrogen atom. A possible family of two-particle optically-inactive bound states}}\\

\vspace{.5cm}

A.I. Agafonov\\

\vspace{.5cm}

National Research Center "Kurchatov Institute", Moscow 123182, Russian Federation\\

Moscow Aviation Institute (National Research University), Moscow, 125993, Russian Federation\\

\vspace{.5cm}

Agafonov\_AIV@nrcki.ru, aiagafonov7@gmail.com

\end{center}

\par {\bf Highlights}\\
\begin{itemize}
\item{The invariant mass of free particles is used to derive a bound-state equation for hydrogen atom}
\item{This bound-state equation has the well-known solutions for the hydrogen single-particle states}  
\item{This bound-state equation has a new solutions for two-particle bound states}
\item{In two-particle bound states, the electron and proton have different scales of internal motions}
\item{Radiative processes with the two-particle state should be two-photon ones suppressed by proton mass.}
\end{itemize}

\vspace{.5cm}

\abstract{The invariant mass of free particles is used to derive a bound-state equation for the hydrogen atom at rest. This equation has the 
well-known solutions for the single-particle states. Existence of two-particle bound states, for which the electron and proton have different 
scales of intraatomic motions, is predicted. Two two-particle states are obtained numerically and discussed in detail. 
Radiative processes involving the two-particle bound state, should result in simultaneous changing the electron and the proton states and 
occur through the simultaneous two-photon emission. These processes will be extremely improbable, in comparison with electronic two-photon 
transitions, since they will be additionally suppressed by the large proton mass.}

\vspace{.3cm}


\section{Introduction}
The hydrogen atom is one of the most studied quantum objects. At present, the optical frequencies of the atom have been measured
with very high accuracy \cite{bib1}. It gives a unique opportunity to use the spectroscopy methods for the experimental 
verification of atomic level theories. 

\par Both the non-relativistic and relativistic theories for hydrogen bound states are well known. These theories have one thing in common:
the hydrogen atom, consisting of the two interacting particles, is described by the single-particle bound states.
	
\par In non-relativistic quantum mechanics, the solution of the Schrödinger equation for the atom is found using the method
of variable separation \cite{bib2}. Then, in the center-of-mass frame, the problem is reduced to the motion of a particle with 
the electron charge and the reduced mass in the Coulomb field. The eigenstates of the transformed equation are the single-particle states.

\par In quantum electrodynamics, bound two-body systems are studied using the Bethe-Salpeter equation \cite{bib3,bib4}. 
Its exact solution for hydrogen is difficult to find even in the ladder approximation for the interaction function \cite{bib5}. 
Therefore, the electron motion in the external field of the fixed proton is investigated \cite{bib4}.  
In the frame of reference associated with a fixed proton, the bipinor wave function of the electron depends only on its radius-vector.
The single-particle bound states for the Dirac equation with the external field are considered as the hydrogen states. 

\par It is usually accepted that in the Hamiltonian $H = H_{0} + V$, $H_{0}$ is the operator of the energy of free, non-interacting 
particles, and $V$ is the interaction between them. In the works \cite{bib6,bib7,bib8,bib9,bib10} a system of two particles with the same mass 
was studied. The operator $H_{0}$ was represented as the sum of the relativistic energies of free particles whose momenta are equal 
in absolute value. As a result, the bound-state spectrum of the Hamiltonian $2\sqrt{p^2+m^2}+V$ was studied. That is, the two-particle problem
 is again reduced to the one-particle problem.

\par The well-known expansion of the function $\frac{1}{\vert {\bf r}_{e}-{\bf r}_{p}\vert}$ (${\bf r}_{e}$ and ${\bf r}_{p}$ are 
the radius-vectors of the electron and the proton, respectively) in spherical functions \cite{bib11} leads us to the following reasoning: 
in the reference frame in which the atom is at rest, the hydrogen atom could be represented by bound states which depend on 
the absolute values of the particles radius-vectors and the angle between them. These states are the two-particle bound states.
They can be called electron-proton bound states.

\par In the present paper, we propose a relativistic-kinematic method to study the hydrogen bound states. The method is based on 
the fact that in relativistic mechanics the total energy and the total momentum of free particles systems are additive quantities, 
but their total mass is not additive \cite{bib12}. However, the mass of the systems is the invariant; it does not change when moving 
from one frame of reference to another. This circumstance makes it possible to use the invariant mass to construct a bound state equation.
This equation allows us to study the two-particle wave functions that take into account the particle mass difference,  which results in
significantly different spatial scales of the internal motion of particles. Two wave functions from this family is obtained by numerical methods. 
Radiative processes which involve the electron-proton bound states, are discussed. 

\section{The bound-state equation}
In relativistic mechanics, the total energy and the total momentum of the free particles system are additive \cite{bib12}: 
\begin{equation}\label{1}  
E_{s}=\sqrt{m^2+p^2}+\sqrt{M^2+q^2} 
\end{equation}
and 
\begin{equation}\label{2}  
{\bf P}_{s}={\bf p}+{\bf q}.
\end{equation}
Here $m$ and ${\bf p}$ are the electron mass and momentum, $M$ and ${\bf q}$ are the proton mass and momentum. 

\par The mass of the system is not additive. However, the mass defined as: 
\begin{equation}\label{3}  
m_{s}^2=E_{s}^2-{\bf P}_{s}^2,
\end{equation} 
is invariant in all frames of reference.

\par It is important that the energy of the system at rest is: 
\begin{equation}\label{4}  
H_{0}=m_{s}. 
\end{equation} 

\par Substituting \eqref{3} into \eqref{4} and considering the interaction between the particles, we obtain the Schrödinger-like equation: 
\begin{equation}\label{5}  
\Bigl[\sqrt{(m+M)^2 +2mM\Bigl[\Bigl(1+\frac{{\hat {\bf p}^2}}{m^2}\Bigr)^{1/2}\Bigl(1+\frac{{\hat {\bf q}^2}}{M^2}\Bigr)^{1/2}-1\Bigr]
-2{\bf p}{\bf q}}-\frac{\alpha(1-{\hat {\bf v}_{e}}{\hat {\bf v}_{p}})}{\vert {\bf r}_{e}-{\bf r}_{p} \vert}\Bigr]\psi=E\psi.
\end{equation}
Here the interaction of the particles through the vector potential is taken into account, ${\hat {\bf v}_{e}}$ and ${\hat {\bf v}_{p}}$ 
are the electron and the proton velocity operators. 

\par We present the energy $E={\cal E}+m+M$ with ${\cal E}\propto \alpha^2 m$. In the expansion of the left-hand side of Eq. \eqref{5} we 
restrict ourselves to the term of order of $p^4/m^3$. For the hydrogen states, the interaction through the vector potential can be 
omitted since the velocities of the particles are small compared to the speed of light in vacuum. Then we obtain:
\begin{equation}\label{6}  
\left[\frac{1}{2(m+M)}\Bigl(\sqrt{\frac{M}{m}}{\hat{\bf p}}-\sqrt{\frac{m}{M}}{\hat{\bf q}}\Bigr)^2-\frac{\alpha}
{\vert{\bf r}_{e}-{\bf r}_{p}\vert} -\frac{M^2 (M+3m) {\hat{\bf p}}^4}{8m^3 (M+m)^3} \right] \psi({\bf r}_{e},{\bf r}_{p})=
{\cal E}\psi.
\end{equation} 
 
\par The term proportional to $p^4$ on the left side of Eq. \eqref{7} determines the fine splitting of the energy levels. Given $m<<M$, 
the term is reduced to  $p^4/8m^3$ that leads to the well-known expression for the fine structure. Now we omit this term. Then Eq. 
\eqref{7} is reduced to:
\begin{equation}\label{7}  
\left[\frac{1}{2(m+M)}\Bigl(\sqrt{\frac{M}{m}}{\hat{\bf p}}-\sqrt{\frac{m}{M}}{\hat{\bf q}}\Bigr)^2-\frac{\alpha}
{\vert{\bf r}_{e}-{\bf r}_{p}\vert} \right] \psi({\bf r}_{e},{\bf r}_{p})=
{\cal E}\psi({\bf r}_{e},{\bf r}_{p}).
\end{equation} 

\par One can transform Eq. \eqref{7} in terms of new independent variables:
\begin{equation}\label{8} 
{\bf R}=\frac{m{\bf r}_{e}+M{\bf r}_{p}}{m+M}, \hspace{1cm}  {\bf r}={\bf r}_{e}-{\bf r}_{p},
\end{equation} 
Then, it is easy to obtain that the transformed equation is reduced to the Schrödinger equation:
\begin{equation}\label{9} 
\Bigl(-\frac{\hbar^2}{2\mu}\Delta_{{\bf r}}-\frac{\alpha}{r}\Bigr)\psi({\bf r})={\cal E}\psi({\bf r}).
\end{equation} 
Here $\mu=\frac{mM}{m+M}$ is the reduced mass. 

\par Note that in Eq. \eqref{9}, the term related to the center-of-mass motion, is absent. This is due to the fact that Eq. \eqref{7} were obtained 
in the reference frame in which the atom is at rest. In this case, it makes no sense to use the replacement \eqref{8}. 
Instead, we are looking for solutions of the original equation \eqref{7} which depend on the independent variables ${\bf r}_{e}$ and ${\bf r}_{p}$. 

\section{Bound-state integral equation}
The differential equation \eqref{7} corresponds to the following integral equation:
\begin{equation}\label{10}  
\psi({\bf r}_{e},{\bf r}_{p})=-\alpha \int {\bf r}_{e}^{\prime} \int {\bf r}_{p}^{\prime} \int \frac{d{\bf p}}{(2\pi)^3}
\int \frac{d{\bf q}}{(2\pi)^3} e^{i{\bf p}({\bf r}_{e}-{\bf r}_{e}^{\prime})+i{\bf q}({\bf r}_{p}-{\bf r}_{p}^{\prime})}
G({\bf p},{\bf q};{\cal E})\frac{1}{\vert{\bf r}_{e}^{\prime}-{\bf r}_{p}^{\prime}\vert}\psi({\bf r}_{e}^{\prime},{\bf r}_{p}^{\prime}),
\end{equation} 
where, by analogy with the well-known structure of bound-state integral equations, the function $G$ should be considered as the two-particle propagator in the momentum space:
\begin{equation}\label{11}  
G({\bf p},{\bf q};{\cal E})=\frac{1}{{\cal E}-\frac{1}{2(m+M)}\Bigl(\sqrt{\frac{M}{m}}{\bf p}-\sqrt{\frac{m}{M}}{\bf q}\Bigr)^2}.
\end{equation} 

\par  In the momentum space Eq. \eqref{10} takes the form: 
\begin{equation}\label{12}  
\psi({\bf p},{\bf q})=-\frac{\alpha}{2\pi^2} G({\bf p},{\bf q};{\cal E}) \int \frac{d{\bf k}}{({\bf p}-{\bf k})^2}\psi({\bf k},{\bf p}+{\bf q}-{\bf k})
\end{equation} 

\par In Eq. \eqref{12} the binding energy of the two-particle system is given by:
\begin{equation}\label{13}  
{\cal E}=T+U,
\end{equation} 
where the average kinetic energy is:
\begin{equation}\label{14}  
T=<\psi({\bf p},{\bf q})\vert \frac{1}{2(m+M)}\Bigl(\sqrt{\frac{M}{m}}{\bf p}-\sqrt{\frac{m}{M}}{\bf q}\Bigr)^2 \vert \psi({\bf p},{\bf q})>
\end{equation} 
and the average potential energy is
\begin{equation}\label{15}  
U=-<\psi({\bf p},{\bf q})\vert \frac{\alpha}{2\pi^2} \int \frac{d{\bf k}}{({\bf p}-{\bf k})^2} \vert \psi({\bf k},{\bf p}+{\bf q}-{\bf k})>.
\end{equation} 
\par The wave function is normalized: 
\begin{equation}\label{16}  
<\psi({\bf p},{\bf q})\vert \psi({\bf p},{\bf q})>=1
\end{equation}

In the investigated state, the kinetic energies of the proton and the electron are comparable in order of magnitude. The electron kinetic 
energy must be of the order of $\alpha^2 m$ and, respectively, the average momentum of the electron is $<p>\propto \alpha m$. Then, the proton average 
momentum should be of the order of $<q>\propto \alpha \sqrt{mM}$. It means that in the supposed state, $<q>>><p>$. In the stationary steady state, the 
average momenta of the particles must be zero, $<{\bf p}>=0$ and $<{\bf q}>=0$. 

\par The value $T$ given by \eqref{14}, should be considered as the total kinetic energy of the state. Attention is drawn to the symmetry of the equation 
with respect to particle masses. But this symmetry leads to a surprising result: the proton moment does almost not contribute to the electron-proton 
state energy. Indeed, for $p\simeq q\sqrt{m/M}$ we have $p>>q\frac{m}{M}$, and in Eq. \eqref{14} the term $\sqrt{\frac{m}{M}}{\bf q}$ can be omitted.

\par Below, we calculate the energy using the original expression \eqref{14} having the three terms. Their values will be given.
Despite the proton large momentum, the absence of the proton contribution to kinetic energy helps to the confinement of the particles in the 
electron-proton bound state. Of course, for this state, the virial theorem does not hold.

\section{Numerical solution of Eq. \eqref{12}}
In the investigated state, the wave function depends on the absolute values of the particles momenta and the angle between them. $\psi(p,q,\theta)$.
The energy of the state is ${\cal E}=-\alpha^2m*\gamma$, where $\gamma>0$ is the numerical factor. We introduce $\eta=\sqrt{\frac{m}{M}}$ 
and dimensionless variables: $p=\alpha m x$, $q=\alpha \sqrt{mM}y$ and $k=\alpha mz$. Then Eq. \eqref{12} for the wave function is rewritten as:
\begin{equation}\label{17}  
\psi(x,y,\theta)=\frac{1}{\pi^2}\frac{1}{2\gamma +\frac{1}{1+\eta^2}({\bf x}-\eta {\bf y})^2} \int\frac{d{\bf z}}{({\bf z}-{\bf x})^2}
\psi(z,\vert {\bf y}+\eta {\bf x}-\eta {\bf z}\vert,\theta_{s}). 
\end{equation} 
Here $\theta_{s}$ is the angle between $\bf z$ and ${\bf y}+ \eta {\bf x}-\eta {\bf z}$.

\par Eq. \eqref{13} for the energy in units $\alpha^2 m$ is rewritten as: 
\begin{displaymath}
\gamma=-\frac{1}{2(1+\eta^2)}<\psi(x,y,\theta)\vert (x^2-2\eta xy cos \theta+\eta^2 y^2)\vert \psi(x,y,\theta)>+ 
\end{displaymath}
\begin{equation}\label{18}  
\frac{1}{2}<\psi(x,y,\theta)\vert \int\frac{d{\bf z}}{({\bf z}-{\bf x})^2} \psi(z,\vert {\bf y}+\eta {\bf x}- \eta {\bf z}\vert,\theta_{s})>. 
\end{equation} 

\par Without loss of generality, we assume that the vector ${\bf y}+\eta {\bf x}$ is directed along the $z-$axis.  Then Eq. \eqref{17} takes the form:
\begin{displaymath}
\psi(x,y,\theta)=\frac{2}{\pi}\frac{1}{2\gamma+\frac{1}{1+\eta^2}(x^2-2\eta xy cos \theta+\eta^2 y^2)}\int_{0}^{\infty}z^2dz 
\end{displaymath}
\begin{equation}\label{19}  
\int_{0}^{\pi} \frac{\psi(z,y_{s},\theta_{s}) \sin \theta_{z} d\theta_{z}}
{\sqrt{(z^2+x^2-2xz\cos \theta_{z}\cos \theta_{x})^2-4x^2 z^2 \sin^2 \theta_{z}\sin^2 \theta_{x}}}. 
\end{equation} 
Here the following functions are introduced:
\begin{equation}\label{20}  
y_{s}=\sqrt{D^2-2\eta Dz\cos \theta_{z}+\eta^2 z^2}
\end{equation} 
and
\begin{equation}\label{21}
\cos \theta_{s}=\frac{D\cos \theta_{z}-\eta z}{y_{s}}.
\end{equation}   
The angle $\theta_{z}$ is the polar angle of the vector ${\bf z}$, and the function $D=\vert {\bf q}+{\bf p} \vert $ has the form:
\begin{equation}\label{22}  
D=\sqrt{y^2+2\eta xy\cos \theta + \eta^2 x^2}
\end{equation} 
in the dimensionless variables.

The angle $\theta_{x}$ is the polar angle of the vector ${\bf p}$.  This angle is related to the angle $\theta$ as:
\begin{equation}\label{23}
\sin \theta_{x}=\frac{y\sin \theta}{D}.
\end{equation}   

\section{Numerical results}
An iteration method is used for solving Eq. \eqref{19}. The function $\psi(x,y,\theta)$ is represented by a matrix $\psi_{m}(i,j,k)$ 
of the dimension $101\times101\times121$. Here $m$ is the iteration number. An important point is the choice of the initial matrix $\psi_{0}(i,j,k)$. 
Below we present the results for two different choices. First, we used:
\begin{equation}\label{24}
\psi_{0}(i,j,k)=\frac{1}{(1+a_{1}x(i)^2)^2(1+a_{2}y(j)^2)^2},
\end{equation}
where $a_{1}<a_{2}$. The matrix \eqref{24} has no zeros, and this choice is related to the $1s$ momentum-space wave function. After calculating the 
matrix $\psi_{1}(i,j,k)$, the function for the next iteration is $\psi_{m+1}(i,j,k)=\xi \psi_{m}(i,j,k)+\sqrt{1-\xi^2}\psi_{m-1}(i,j,k)$. 
Here $m=1,2,...N_{it}$, $\xi=0.97$.

\begin{figure}[h]
\centering
\includegraphics[width=10cm]{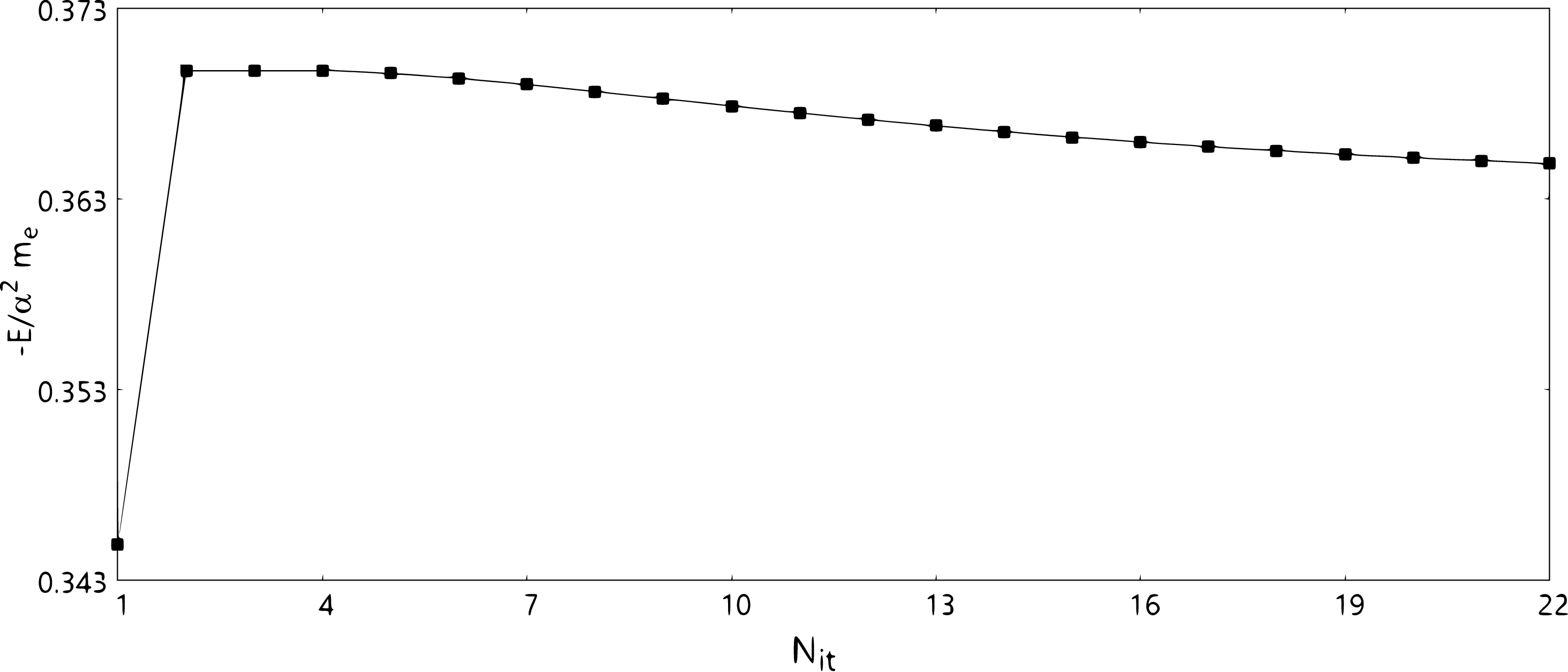}
\caption{The iteration dependence of the energy of the electron-proton bound state.
\label{f1}}
\end{figure}

\par To demonstrate the convergence of the iterative procedure, Fig. 1 shows the iteration dependence of the energy \eqref{18}. It is clearly seen 
that the energy tends to a certain value with increasing the iteration number. This value is ${\cal E}=-0.364\alpha^2 m_{e}$. Similar curves were 
obtained for the iteration dependence of the average values of $<p^2>$ and $<q^2>$. Their asymptotic values are $<p^{2}>=0.383 \alpha^2 m^2$ and  
$<q^{2}>=0.204 \alpha^2 mM$. Hence, the electron average kinetic energy $T_{e}=0.191\alpha^2 m$ and that for the proton $T_{p}=0.102\alpha^2 m$. 
But, as discussed above, this proton energy practically does not contribute to the energy \eqref{18}. The motion of the proton leads to the two 
contributions: $E_{ep}=-\frac{\eta}{1+\eta^2} <xy\cos(\theta)>$ and $E_{p}=\frac{\eta^2}{2(1+\eta^2)}<y^2>$. In the usual units, 
$E_{ep}=-0.68\times 10^{-3}\alpha^2 m$ and $E_{p}=0.56\times 10^{-4}\alpha^2 m$. 

\par The average potential energy $U$  given by Eq. \eqref{15}, is $U=-0.556 \alpha^2 m_{e}$. It means that 
\begin{equation}\label{25}
<\psi({\bf r}_{p},{\bf r}_{p})\vert \frac{1}{\vert {\bf r}_{e}-{\bf r}_{p} \vert }\vert \psi({\bf r}_{e},{\bf r}_{p})> =0.556 a_{B}^{-1}.
\end{equation}
This value is intermediate between $<1s\vert r^{-1} \vert 1s>=a_{B}^{-1}$ and  $<2s\vert r^{-1} \vert 2s>=0.25a_{B}^{-1}$.

\begin{figure}[h]
\centering
\includegraphics[width=10cm]{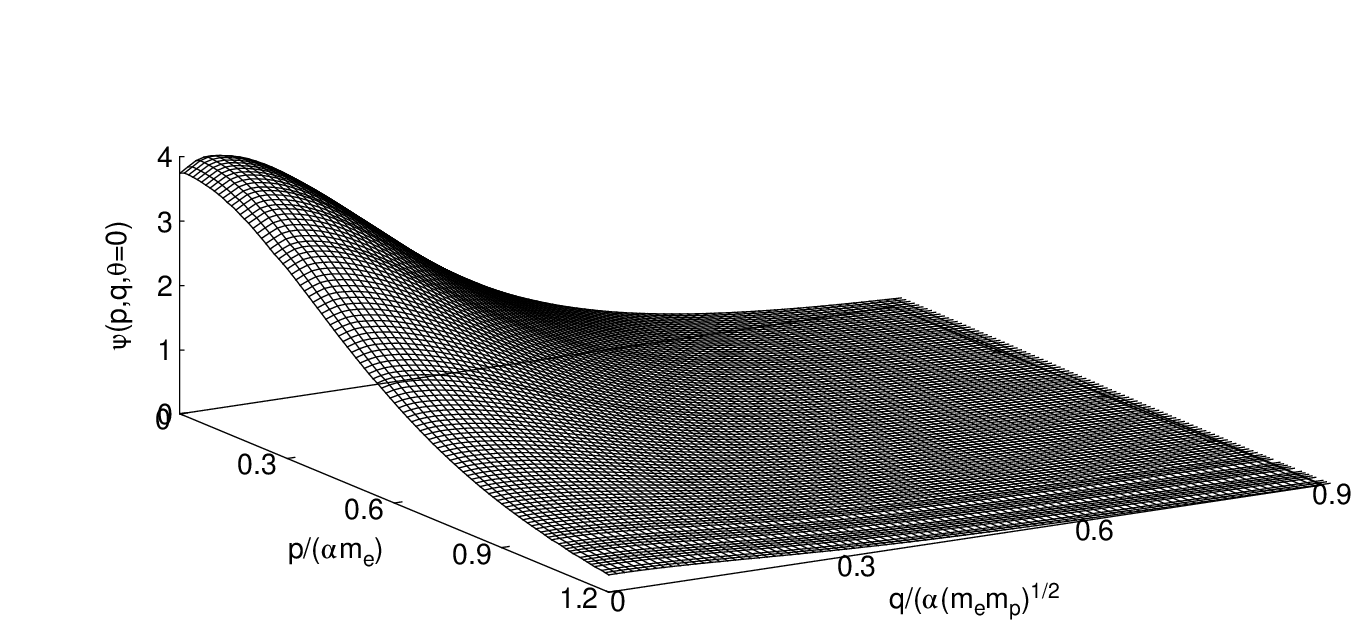}
\caption{The wave function $\psi(x,y,\theta=0)$ satisfying Eq. \eqref{19} with the initial matrix \eqref{24}.
\label{f2}}
\end{figure}

\par It turned out that when choosing \eqref{24}, the calculated function $\psi(x,y,\theta)$ depends weakly on the angle $\theta$. 
The wave function for the angle $\theta=0$ is shown in Fig. 2. It have no zeros, just like the ground state wave function of the hydrogen atom. 
The energy of this electron-proton state ${\cal E}=-0.364\alpha^2m$ is greater than the ground state energy ${\cal E}_{1s}=-0.5\alpha^2\mu$, but is less than the energy of the first excited state ${\cal E}_{2s}=-0.125\alpha^2\mu$.   

\par There are significant differences between the momentum space wave function shown in Fig. 2, and the $1s$ wave function. In the center of mass system, 
${\bf p}+{\bf q}=0$ for the well-known bound states, and the average momenta of the particles are the same. For the $1s$ state $<p>=<q>=0.849\alpha m$. 
Therefore, the proton kinetic energy is 1836 times less than that for the electron.

\par For the found state, the sum ${\bf p}+{\bf q}$ is uncertain, $<{\bf p}>=0$ and $<{\bf p}>=0$, and the average values $<p>=0.566\alpha m$ and 
$<q>=0.381\alpha \sqrt{mM}$. Hence, the kinetic energies of the proton and the electron are of the same order.

\begin{figure}[h]
\centering
\includegraphics[width=10cm]{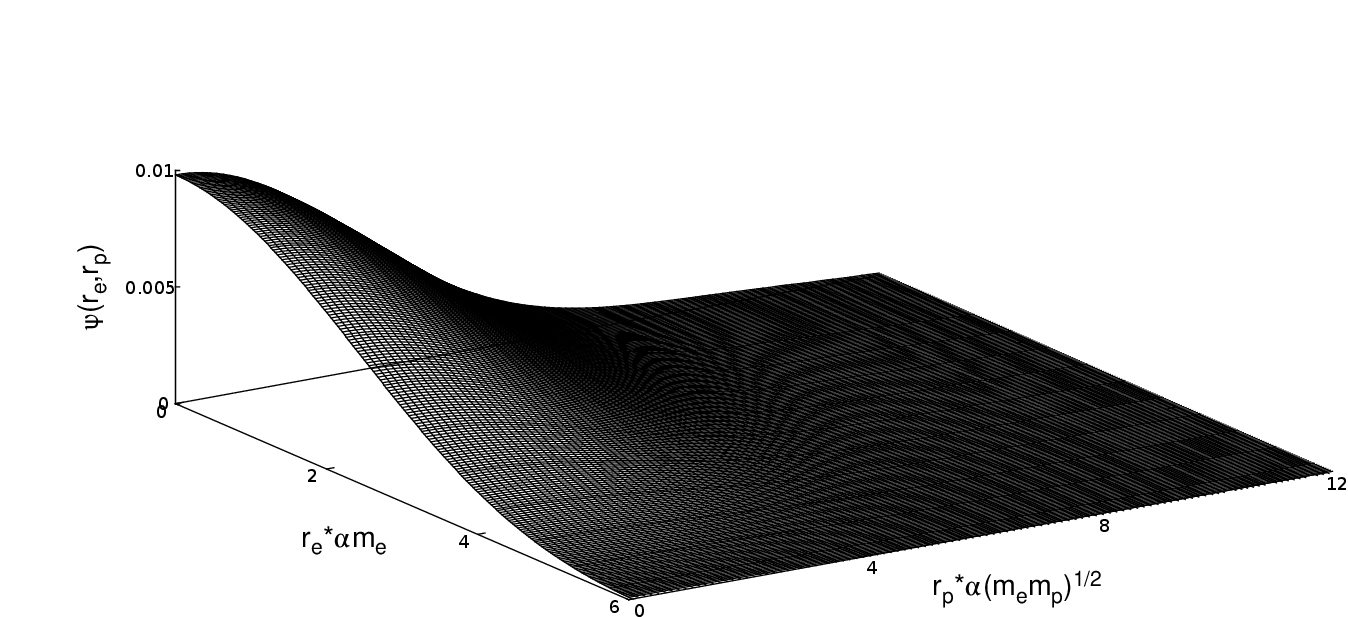}
\caption{Fourier transform function of the wave function $\psi(p,q, \theta=0)$ shown in Fig. 2.
\label{f3}}
\end{figure}

\par The wave function $\psi(x,y,\theta)$ depends on the angle $\theta$. In order to elucidate this dependence, we calculate the electron kinetic 
energy $T_{e}(\theta)$ for the given angle. We obtain that $T_{e}(0)=0.201\alpha^2 m$, $T_{e}(\frac{\pi}{2})=0.184 \alpha^2 m$ and 
$T_{e}(\pi)=0.189\alpha^2 m$ with the angle-averaged value $T_{e}(0)=0.191\alpha^2 m$. Overall, the relative change of the kinetic energy is about 
5 percent. So, the angle dependence of the wave function is relatively weak.

\par The coordinate-space wave function $\psi(r_{e},r_{p})$ which was obtained by the Fourier transform of the function $\psi(p,q, \theta=0)$,    
is demonstrated in Fig. 3.  According to this function, the electron and the proton have significantly different scales of intraatomic motions.
We calculated the average radius-vectors of the electron and the proton: $<r_{e}>=2.45 a_{B}$ and $<r_{p}>=9.805 \sqrt{\frac{m}{M}} a_{B}=0.229 a_{B}$.  
These scales of internal motions differ by an order of magnitude.

\par Now we present numerical results for the second choice of the initial matrix. It has multiple zeros:
\begin{equation}\label{26}
\psi_{0}(i,j,k)=\cos(b_{1}x(i)-b_{2}y(i))/((1+a_{1}x(i)^2)^2(1+a_{2}y(j)^2)^2)
\end{equation}
with constants $b_{1}<b_{2}$. 

\begin{figure}[h]
\centering
\includegraphics[width=10cm]{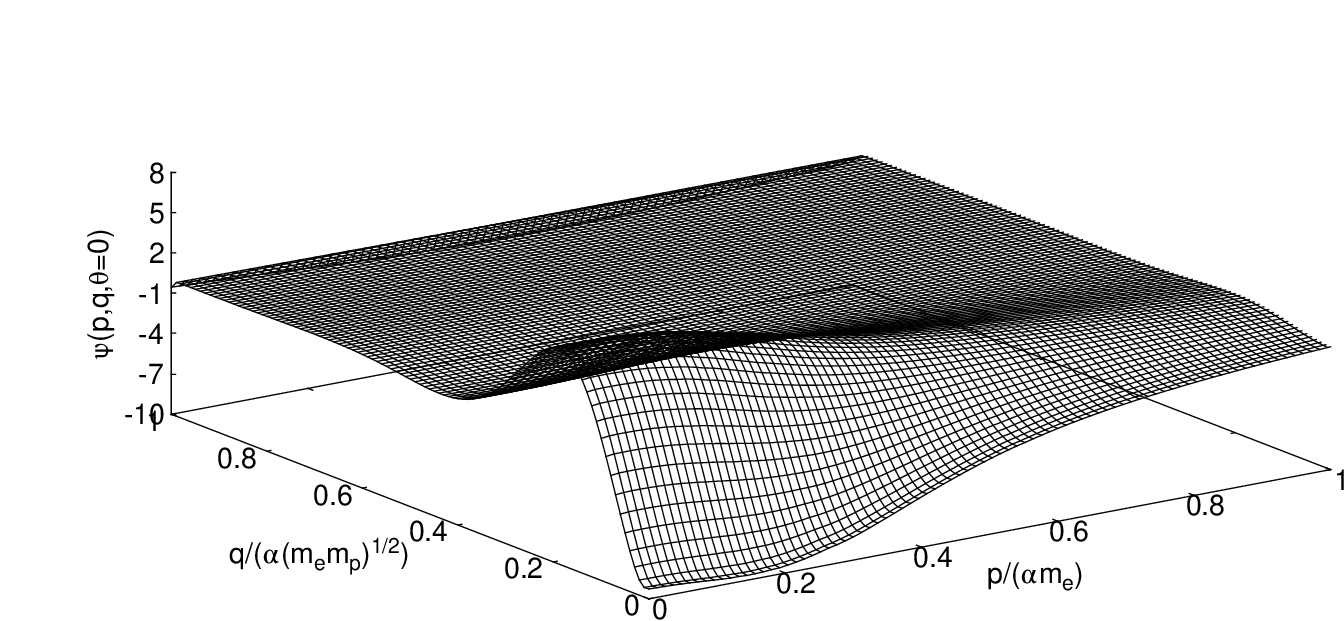}
\caption{The momentum space wave function satisfying Eq. \eqref{16} with the initial condition given by the matrix \eqref{26}.
\label{f2}}
\end{figure}

\par The course of the iterative procedure was similar to that discussed above. The momentum space wave function is shown in Fig. 4. 
It has zeros, and, hence, this state is an excited one. The energy of this state is ${cal E}=-0.329\alpha^2m$, which is greater than that 
for the state shown in Fig. 2. It was obtained: $<p^{2}>=0.280 \alpha^2 m^2$ and $<q^{2}>=0.156 \alpha^2 mM$. Accordingly, the average 
kinetic energy of the electron $T_{e}=0.140\alpha^2 m$ and that for the proton $0.078\alpha^2 m$. The latter does not make a 
significant contribution to ${\cal E}$ for the reason discussed above.

\par The average potential energy $U$ given by Eq. \eqref{15}, is $U=-0.469 \alpha^2 m$. It means that 
\begin{equation}\label{27}
<\psi({\bf r}_{e},{\bf r}_{p})\vert \frac{1}{\vert {\bf r}_{e}-{\bf r}_{p} \vert }\vert \psi({\bf r}_{e},{\bf r}_{p})>=0.469 a_{B}^{-1}.
\end{equation}
Thus, for this two-particle state, the internal motion of particles takes place in a larger space region than that given by Eq. \eqref{25}.
\begin{figure}[h]
\centering
\includegraphics[width=10cm]{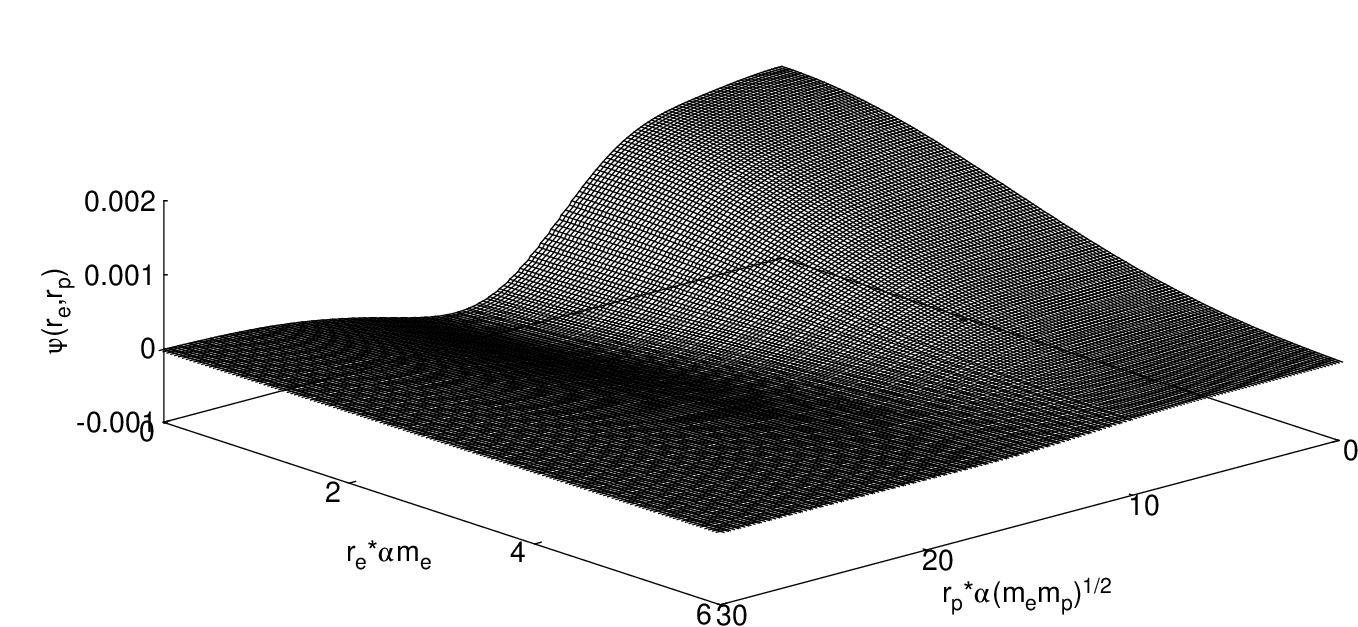}
\caption{The coordinate-space wave function which correspond to the function shown in Fig. 4.
\label{f3}}
\end{figure}

\par The coordinate-space wave function $\psi(r_{e},r_{p})$ was obtained by the Fourier transform of the function shown in Fig. 4. 
It is demonstrated in Fig. 5.  According to this function, the electron and the proton have significantly different scales of intraatomic motions.
We calculated the average radius-vectors of the electron and the proton: $<r_{e}>=2.89 a_{B}$ and $<r_{p}>=0.267 a_{B}$.  

\section{Conclusion: Optically-inactive bound states}
In conclusion, we would like to discuss the following: if in the hydrogen atom, the two-particle states obtained above, exist, then 
why their did not detected earlier? In this regard, the most important are spectroscopic studies.

\par Regardless of whether a two-particle bound state is the final or initial state, any optical process must be accompanied by a change in the 
internal motions of both the electron and the proton. That is, during a radiative transition, both the electron and the proton change 
simultaneous their states. But these particle states are not independent. They can only change simultaneously because they are unique determined 
by the two-particle bound state.

\par Therefore, any radiative process involving any two-particle bound state, can be expected to be a transition with simultaneous emission 
of two photons. One of them is emitted by the electron, and the second is emitted by the proton. It should be noted that in the hydrogen atom, 
the most probable process of the $2s\to 1s$ transition is the transition with the simultaneous emission of two photons by the electron 
(see \cite{bib4} and referenced therein). However, the probability of this process, is quite small compared to the single-photon 
electric-dipole transitions. The process probability is $8s^{-1}$. The radiative process under discussion should have a lower probability 
by many orders of magnitude, since the two-photon transition will be suppressed by the large proton mass. Therefore, these two-particle states
 can be optically inactive. 

\par Note also that the relativistic-kinematic approach presented above is easily generalized to systems of many particles with pair interactions 
between them. In this case, the generalization of equation \eqref(5) to N particles is:
\begin{equation}\label{28}
\left[ \sqrt{\Bigl(\sum_{i=1}^{N} {\hat E}_{i}\Bigr)^2 - \Bigl(\sum_{i=1}^{N} {\hat {\bf p}}_{i}\Bigr)^2}-
\frac{1}{2}\sum_{i}^{N}\sum_{j\neq i}^{N}V_{ij}\right]\psi(\{{\bf r}_{i}\})=
E\psi(\{{\bf r}_{i}\}),
\end{equation}
where ${\hat E}_{i}=\sqrt{m_{i}^2+{\hat {\bf p}}_{i}^2}$ is the operator of the $i$-particle energy, $m_{i}$ and ${\hat {\bf p}}_{i}$ its mass 
and momentum operator. The only constraint on particle momenta in $N$-particle bound state is
\begin{equation}\label{29}
<\psi(\{{\bf r}_{j}\})\vert {\hat {\bf p}}_{i=1,\cdot, N} \vert \psi(\{{\bf r}_{j}\})>=0.
\end{equation}

\par {\bf CRediT authorship contribution statement}\\
This article has one author.

\par {\bf Declaration of Competing Interest}\\
The authors declare that they have no known competing financial interests or personal relationships that could have appeared to influence the work reported in this paper.

\par {\bf \bf Data availability}\\
No data was used for the research described in the article.


\begin{thebibliography}{00}
\bibitem{bib1} Ch.G. Parthey, A. Matveev, J. Alnis, et al., Phys. Rev. Lett. 107, 203001 (2011).
\bibitem{bib2} L.D. Landau and E.M. Lifshitz, Quantum mechanics. Non-relativistic theory, Pergamon Press, 1965.
\bibitem{bib3} H.A. Bethe and E.E. Salpeter, A relativistic equation for bound-state problems, Phys. Rev. A 84, 1232 (1951).
\bibitem{bib4} V B Berestetskii, L. P. Pitaevskii, E.M. Lifshitz, Quantum electrodynamics, Elsevier, 2012. 
\bibitem{bib5} H.A. Bethe and E.E. Salpeter, Quantum Mechanics of One- and Two-Electron Atoms, Springer, 1957.
\bibitem{bib6} E.E. Salpeter, Phys. Rev. 87, 328 (1952).
\bibitem{bib7} I.W. Herbst, Commun.Math. Phys. 53, 285 (1977).
\bibitem{bib8} C. Semay, An upper bound for asymmetrical spinless Salpeter equations, Phys.Lett. A 376 2217 (2012).
\bibitem{bib9} W. Lucha, F.F. Schöberl, Int. J. Mod. Phys. A 34, 1950028 (2019).
\bibitem{bib10} M.N. Sergeenko, arXiv:1912.07598v1 [hep-ph].
\bibitem{bib11} Handbook of Mathematical Functions: with Formulas, Graphs, and Mathematical Tables. 
                Edited by M. Abramowitz and I.A. Stegun, Dover Books on Mathematics, 1046 pages.
\bibitem{bib12} L.B. Okun’, The concept of mass (mass, energy, relativity), Sov. Phys. Usp. 32 629–638 (1989)
\end{thebibliography}
\end{document}